\documentclass[a4paper,11pt]{article}
\pdfoutput=1 

\usepackage{jinstpub} 

\usepackage[load-configurations = abbreviations]{siunitx}
\usepackage{subfig}
\usepackage{supertabular,booktabs}
\usepackage[table,xcdraw]{xcolor}
\usepackage{adjustbox}

\newlength{\subcolumnwidth}
\newenvironment{subcolumns}[1][0.45\columnwidth]
 {\valign\bgroup\hsize=#1\setlength{\subcolumnwidth}{\hsize}\vfil##\vfil\cr}
 {\crcr\egroup}
\newcommand{\nextsubcolumn}[1][]{%
  \cr\noalign{\hfill}
  \if\relax\detokenize{#1}\relax\else\hsize=#1\setlength{\subcolumnwidth}{\hsize}\fi
}
\newcommand{\nextsubfigure}{\vfill}
\newcommand{\dhyphen}{\mbox{--}}

\usepackage{lineno}

\title{\boldmath Test beam results on 3D pixel sensors for the CMS Tracker upgrade at the High-Luminosity LHC}

\author[a,1]{C. Lasaosa\note{Corresponding author.}}

\affiliation[a]{IFCA (CSIC-UC), Santander, Spain}

\emailAdd{clara.lasaosa.garcia@cern.ch}

\abstract{The High Luminosity upgrade of the CERN Large Hadron Collider (HL-LHC) requires new high-radiation tolerant silicon pixel sensors for the innermost part of the tracking detector in the CMS experiment. The innermost layer of the tracker, which is as close as $\SI{3}{\cm}$ from the interaction point, will be exposed to a fluence of $\SI{3.4e16}{n_{eq}\cm^{-2}}$ during the high-luminosity operation period. The 3D pixel sensor technology has been proven to be the best option for such a layer in terms of radiation tolerance and low power consumption. An extensive program aiming at 3D pixel sensors has been carried out in the context of the CMS tracker R\&D activities. The sensors have been produced by Fondazione Bruno Kessler (Trento, Italy) and Centro Nacional de Microelectrónica (Barcelona, Spain). They are connected with the CROCv1 readout chip, which is a prototype of the final version. The modules have been tested in beams at CERN and DESY, before and after irradiation up to an equivalent fluence of about $\SI{1.6e16}{n_{eq}\cm^{-2}}$. An overview of the results obtained in the latest beam test experiments is presented, including hit detection efficiency and spatial resolution. The analysis of collected data shows excellent performance, with around 98\% hit detection efficiencies measured after irradiation.}


\collaboration[c]{on behalf of the CMS Tracker Group}

\proceeding{25$^{\text{th}}$ International Workshop on Radiation Imaging Detectors\\
  30 June - 4 July 2024\\
  Lisbon, Portugal}

\begin{document}
\maketitle
\flushbottom

\section{Introduction}
\label{sec:introduction}

The environment of the High-Luminosity Large Hadron Collider (HL-LHC)~\cite{CERN-LHCC-2017-009} will feature a peak instantaneous luminosity of $\SI{7.5e{34}}{cm^{-2}s^{-1}}$ in the ultimate performance scenario, with up to 200 collisions per bunch-crossing on average. This will enable the CMS experiment to collect an integrated luminosity of $\SI{4000}{fb^{-1}}$ over the project lifetime.

The foreseen radiation levels for the unprecedented high-luminosity scenario lead to the need of upgrading the Compact Muon Solenoid (CMS) experiment. In particular, the innermost region, which is called Inner Tracker (IT) and is made of silicon pixel modules, will be completely replaced. The layer closest to the interaction point will be exposed to a fluence of $\SI{3.4e16}{n_{eq}cm^{-2}}$, and a total ionising dose of $\SI{1.9}{Grad}$ will be accumulated during the high-luminosity operation period. A replacement of this layer is planned after approximately six years of operation, at which point it is estimated that a fluence of $\SI{1.8e16}{n_{eq}cm^{-2}}$ will be reached.

\begin{figure}[hbt!]
\centering
\includegraphics[width=.7\textwidth]{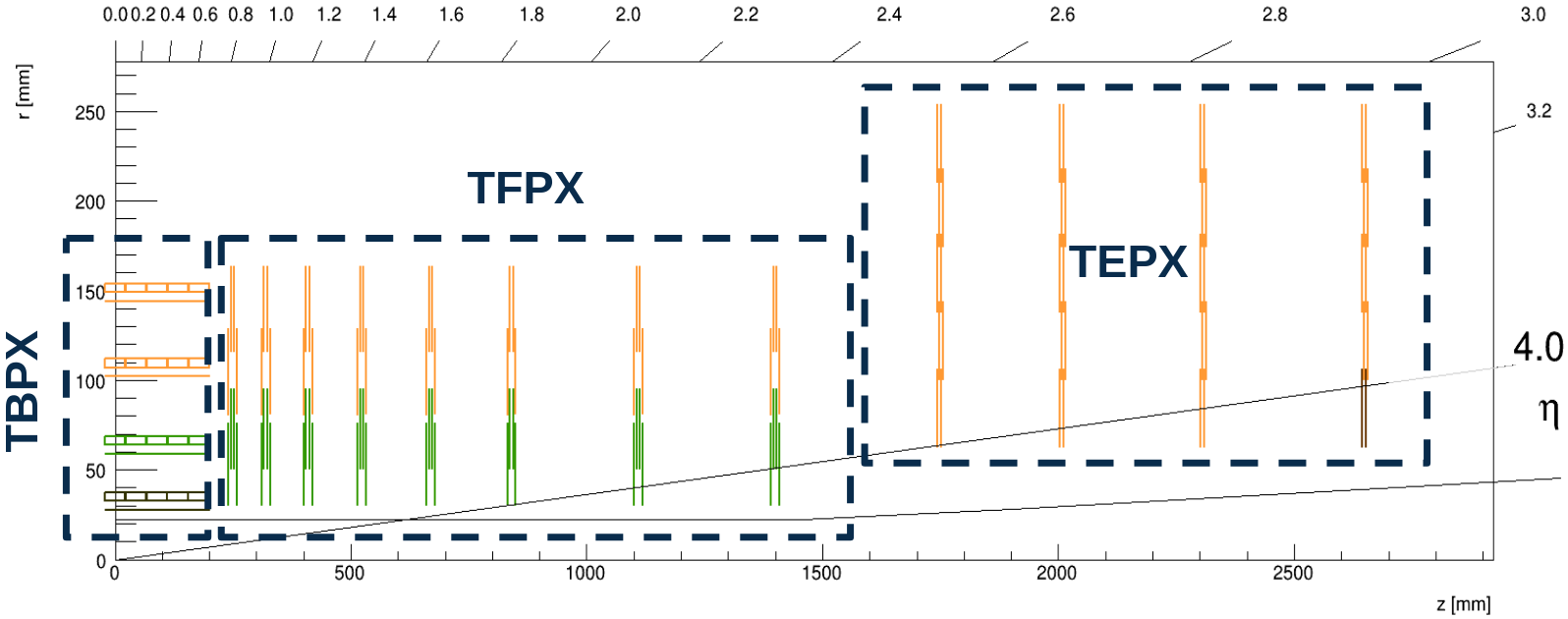}
\caption{One quarter of the upgraded IT layout in the longitudinal view. Planar pixel modules with two and four ROCs are depicted in orange and green, respectively. 3D pixel modules with two ROCs are represented in black.}
\label{fig:ITlayout}
\end{figure}

The upgraded CMS IT layout includes an extension of the tracking coverage, optimization of the layer arrangement and reduction of the material budget, among others. It will consist of three substructures as shown in Fig.~\ref{fig:ITlayout}: Tracker Barrel Pixel (TBPX) with four central layers, Tracker Forward Pixel (TFPX) with eight small double discs at each end and Tracker Extended Pixel (TEPX) with four large double discs per side. It will feature hybrid modules with two and four readout chips (ROCs) and it is designed to allow for module replacement if needed.

High radiation tolerance, low power dissipation, increased granularity for improved spatial resolution, and high hit detection efficiency have been crucial requirements in sensor optimization due to the more demanding operational conditions of the HL-LHC. Consequently, the baseline design consists of n$\dhyphen$in$\dhyphen$p silicon pixel sensors with an active thickness and a pixel cell size reduced to $\SI{150}{\mu m}$ and $25\times100$~$\SI{}{\mu m^{2}}$, respectively. For sensor technology, 3D pixels have been selected as the optimal choice for the innermost layer of the TBPX, while planar pixels will be employed elsewhere. This selection is based on detailed simulations that indicate thermal runaway issues with planar sensors in the innermost layer.

\section{Hybrid 3D pixel silicon modules}
\label{sec:modules_description}

\begin{figure}[hbt!]
\centering
\begin{subcolumns}[0.55\textwidth]
  \subfloat[]{\includegraphics[width=8cm]{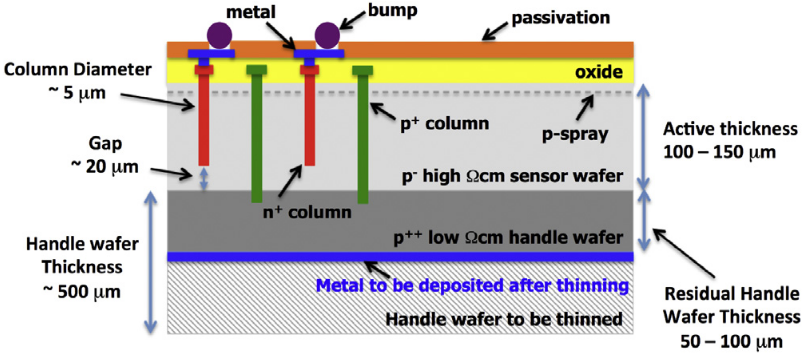}
  \label{fig:FBK_3D_design}}
\nextsubcolumn[0.43\textwidth]
  \subfloat[]{\includegraphics[width=5.6cm]{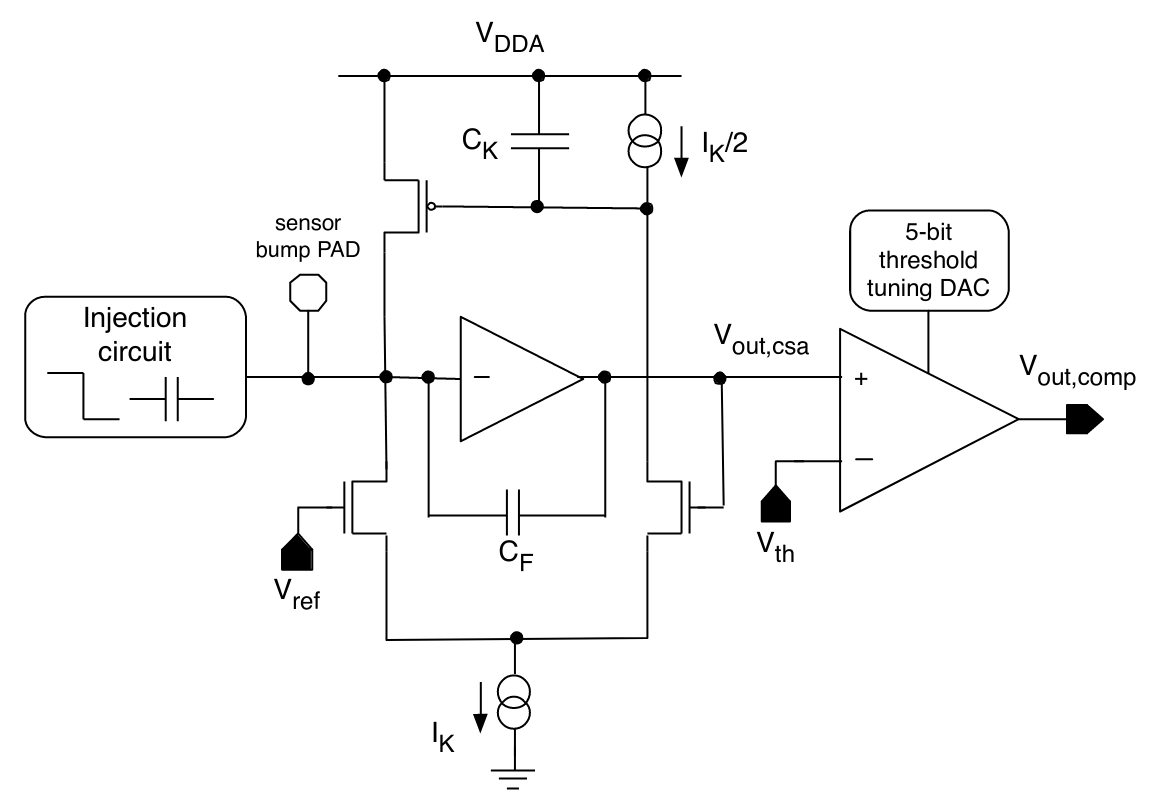}\label{fig:CROC_schema}}
\end{subcolumns}
\vspace{-0.1cm}
\caption{(a) FBK 3D pixel sensor layout, and (b) schematic of the CMS analog linear front-end.}
\end{figure}

3D pixel sensors~\cite{PARKER1997328} consist of cylindrical electrodes that penetrate the bulk perpendicularly as shown in Fig.~\ref{fig:FBK_3D_design}. This design decouples the thickness from the inter-electrode distance, resulting in a lower depletion voltage which reduces the power consumption. Moreover, the shorter collection path for charge carriers leads to increased charge collection efficiency after irradiation.

There have been two manufacturers involved in the R\&D program of the 3D pixel technology for the high-luminosity upgrade of the CMS Tracker: Fondazione Bruno Kessler (FBK)~\cite{FBK} along with Instituto Nazionale di Fisica Nucleare (INFN) in Italy, and Centro Nacional de Microelectrónica (CNM)~\cite{CNM} in Spain. The main differences between the devices of both foundries are the column diameter, which is $\SI{5}{\mu m}$ in FBK and $\SI{8}{\mu m}$ in CNM, as well as the $n^{+}$-column length, which is approximately $\SI{115}{\mu m}$ in FBK and $\SI{130}{\mu m}$ in CNM.

In hybrid modules, the sensor is connected to the front-end electronics through the bump-bonding technique. The CMS ROC (CROC)~\cite{Garcia-Sciveres:2665301}, developed by the RD53 Collaboration, is based on $\SI{65}{nm}$ CMOS technology and features an analog linear front-end whose schematic is shown in Fig.~\ref{fig:CROC_schema}. The readout signal passes through a charge-sensitive amplifier with a Krummenacher feedback loop, which provides a linear discharge of the capacitor. The output signal is then compared against a trimmable threshold for digital conversion using a Time-Over-Threshold (ToT) counter. The matrix is arranged in $432\times336$ square pixels with $\SI{50}{\mu m}$-pitch. All the single-chip devices, which have been tested and whose results are shown in the following sections, feature the prototype of the final chip called CROCv1. For successful operation, a set of calibrations is performed to tune the modules to a low threshold while keeping the number of masked pixels minimal.

Table~\ref{tab:DUTsummary} includes a list of the modules with details about the irradiation and test beam campaigns. The irradiation of the 3D pixel modules was performed at Karlsruhe Institute of Technology (KIT) with protons at $\SI{25}{MeV}$ and at the CERN Proton Synchroton (PS) with protons at $\SI{24}{GeV}$. The data taking was carried out at two facilities using different kinds of test beams: at Deutsches Elektronen-Synchrotron (DESY) with an electron/positron beam at $\SI{5.2}{GeV}$ and at the CERN Super Proton Synchroton (SPS) with a pion beam at $\SI{120}{GeV}$. The test beam setup, similar at both facilities, consists of a EUDET-type telescope~\cite{jansen2016performance} with downstream and upstream triplets embedded with Mimosa26 sensors, which feature square pixels with a $\SI{18.4}{\mu m}$-pitch. The system is also equipped with a Trigger Logic Unit (TLU) that provides the trigger signal and a reference pixel module to filter out out-of-time tracks, which might be reconstructed by the telescope planes due to their long integration time. The Devices Under Test (DUTs) are placed within a cold box located between both telescope arms over a stage that allows rotation and translation of the modules to different positions. The cold box keeps the modules at low temperatures, which is crucial after irradiation.
\begin{table}[hbt!]
    \centering
    \caption{Summary of the single-chip 3D pixel devices that have been tested on beam and whose results are shown in the following sections.}
    \begin{adjustbox}{width=1\textwidth,center=\textwidth}
    \begin{tabular}{|l|c|c|c|c|cc|}
   \hline
     \hspace{0.7cm}Module & Sensor & Pixel size & Fluence & Irradiation & \hspace{0.7cm}Test beam area &\\
     \hspace{0.85cm}name & manufacturer & [$\mu m^{2}$] & [$\SI{}{n_{eq}\cdot cm^{-2}}$] & facility & \hspace{-1cm}non-irradiated & \hspace{-1cm}irradiated\\ 
    \hline\hline
     \textbf{CROC\_CNM\_1} & CNM & $50\times50$  & - & - & \hspace{-1cm}SPS & \hspace{-1cm}-\\
     \textbf{CROC\_CNM\_2} & CNM & $25\times100$ & $\SI{1e16}{}$ & PS & \hspace{-1cm}SPS & \hspace{-1cm}SPS\\
     \textbf{CROC\_FBK\_1} & FBK & $25\times100$ & - & - & \hspace{-1cm}SPS & \hspace{-1cm}-\\
     \textbf{CROC\_FBK\_2} & FBK & $25\times100$ & $\SI{1e16}{}$ & PS & \hspace{-1cm}SPS & \hspace{-1cm}SPS\\
     \textbf{CROC\_FBK\_3} & FBK & $25\times100$ & $\SI{1e16}{}$ & PS & \hspace{-1cm}SPS & \hspace{-1cm}SPS\\
     \textbf{CROC\_FBK\_4} & FBK & $25\times100$ & $\SI{1.6e16}{}$ & KIT & \hspace{-1cm}- & \hspace{-1cm}DESY\\
     \hline
    \end{tabular}
    \end{adjustbox}
    \label{tab:DUTsummary}
\end{table}

\vspace{-0.4cm}
\section{Methodology and test beam results of non-irradiated modules}
\label{sec:non-irradiated_modules}

The performance of several non-irradiated modules was investigated using test beam. The main four consist of 3D pixel sensors with $25\times100$~$\SI{}{\mu m^{2}}$ pixel size, from either FBK or CNM productions. The most salient differences between the two manufacturers mentioned in Section~\ref{sec:modules_description} are discussed below. Moreover, results from a CNM sample with $\SI{50}{\mu m}$-pitch are included, as this cell design was under consideration during the early stages of the R\&D program.

The module thresholds were tuned at a temperature of $\SI{-10}{\degree C}$ to an average pixel threshold of either $\SI{1000}{e^{-}}$ or $\SI{1200}{e^{-}}$, depending on the sample. In all devices, the percentage of pixels masked during tuning and before data acquisition was below $\SI{1}{\%}$. Pixels were masked if they were either stuck or noisy. A stuck pixel does not respond to charge injections, while a noisy pixel exhibits a per-pixel occupancy higher than $2\times10^{-5}$ in the absence of injected charge.

\begin{figure}[hbt!]
\centering
\begin{subcolumns}[0.4\textwidth]
  \subfloat[]{\includegraphics[width=8.cm]{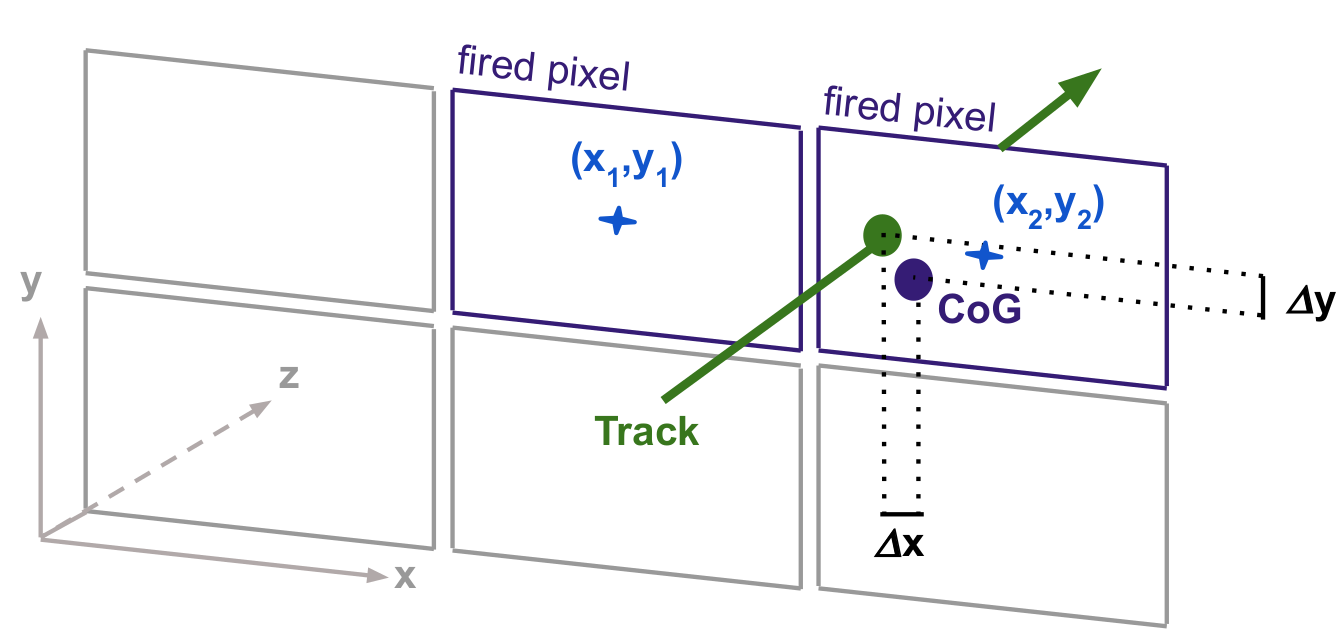}
  \label{fig:cluster_reconstruction}}
\nextsubcolumn[0.4\textwidth]
  \subfloat[]{\includegraphics[width=6.2cm]{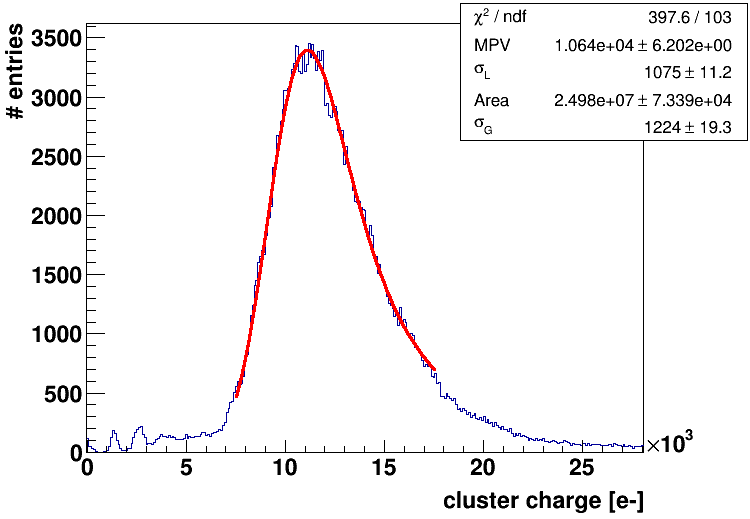}
  \label{fig:clusterCharge_fresh}}
\end{subcolumns}
\vspace{-0.2cm}
\caption{(a) Schematic representation of clustering. The individual hit pixel positions are denoted by $(x_{i},y_{i})$. The track and cluster positions are represented by green and violet filled circles. $\Delta x$ and $\Delta y$ indicate the residuals along both axes. (b) Cluster charge distribution for module CROC\_FBK\_2 fitted to a Landau with most probable value $MPV$ and width $\sigma_{L}$, convoluted with a Gaussian of width $\sigma_{G}$.}
\end{figure}

The clustering algorithm groups adjacent pixels hit by the passage of a single particle into a cluster, estimating the cluster position as the charge-weighted mean of the individual pixel positions (see Fig.~\ref{fig:cluster_reconstruction}). The cluster size corresponds to the number of hit pixels, while the cluster charge is determined by the sum of the individual pixel charges. Figure~\ref{fig:clusterCharge_fresh} shows the cluster charge distribution of module CROC\_FBK\_2 above full depletion ($\SI{30}{V}$) at normal beam incidence. The distribution is fitted to a Landau distribution convoluted with a Gaussian, and the most probable value (MPV) is approximately $\SI{11000}{e^{-}}$, consistent with the expectation from simulations.
\vspace{-0.3cm}

\begin{figure}[hbt!]
\centering
\begin{subcolumns}[0.55\textwidth]
  \subfloat[]{\includegraphics[width=8.2cm]{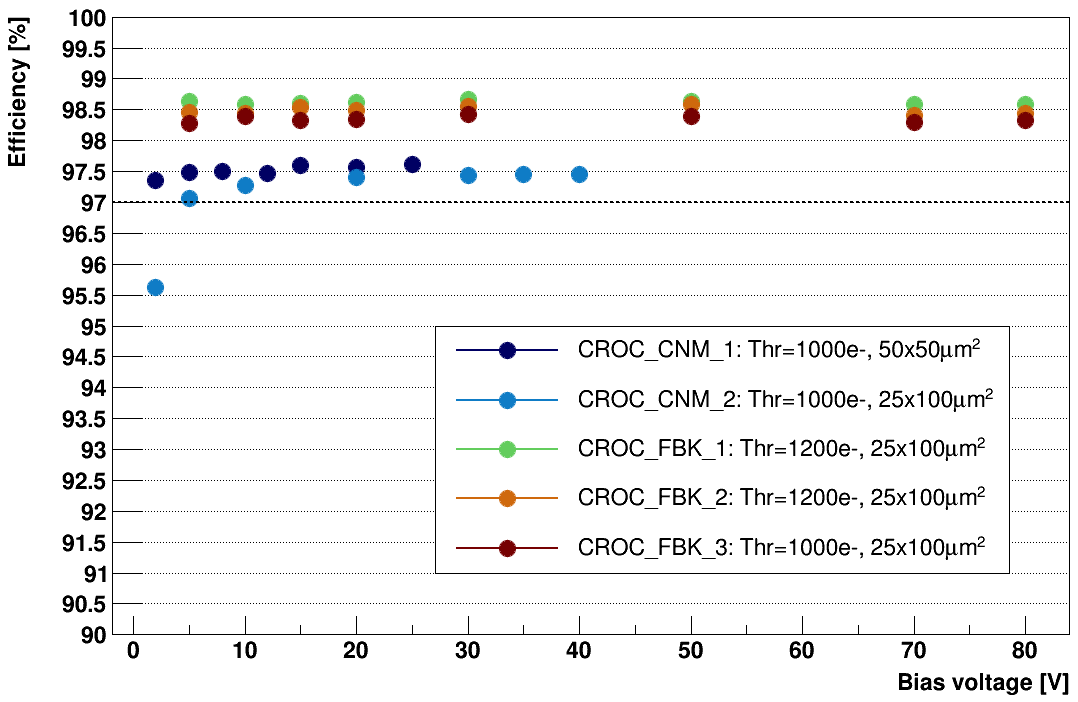}
  \label{fig:biasScan_fresh}}
\nextsubcolumn[0.4\textwidth]
  \subfloat[]{\includegraphics[width=5.9cm]{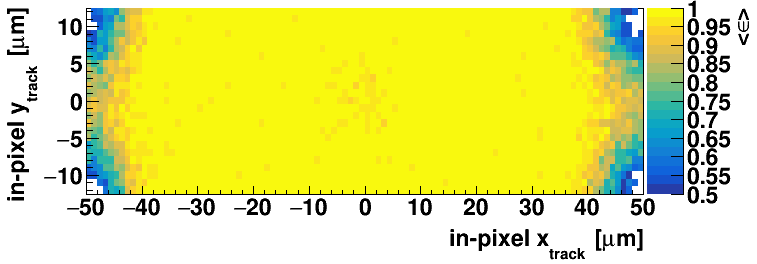}
  \label{fig:eff_map_fresh2V}}
  \nextsubfigure
  \subfloat[][]{\includegraphics[width=5.9cm]{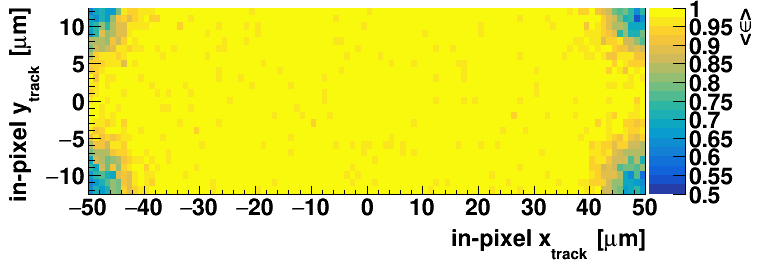}
  \label{fig:eff_map_fresh30V}}
\end{subcolumns}
\vspace{-0.1cm}
\caption{(a) Hit detection efficiency as a function of bias voltage for several non-irradiated modules. 97\% efficiency is indicated by a dashed line. Pixel cell maps of efficiency for module CROC\_CNM\_2 at a bias voltage of (b) $\SI{2}{V}$ and (c) $\SI{30}{V}$.}
\end{figure}

The hit detection efficiency is defined as the ratio of the number of tracks with an associated hit on the DUT to the total number of tracks. For tracks to be considered, they must have an associated hit on all telescope planes and on the reference module. Additionally, a bunch-crossing restriction and a track isolation criterion are applied to the reference module to ensure high-quality tracks. Figure~\ref{fig:biasScan_fresh} shows efficiency above $\SI{97}{\%}$ at normal beam incidence after full depletion, which happens at a bias voltage of $\SI{5}{V}$ or even lower in some modules. The higher efficiency of FBK sensors compared to CNM sensors is attributed to the smaller radius of their electrodes, which are made of passive material. Figures~\ref{fig:eff_map_fresh2V} and~\ref{fig:eff_map_fresh30V} show the efficiency cell maps for module CROC\_CNM\_2 at bias voltages below ($\SI{2}{V}$) and above ($\SI{30}{V}$) full depletion, respectively. These maps illustrate the progression of depletion from the central n$^{+}$-column towards the corners of the pixel cell, where the ohmic columns are located.

\begin{figure}[hbt!]
\centering
\begin{subcolumns}[0.55\textwidth]
  \subfloat[]{\includegraphics[width=8.2cm]{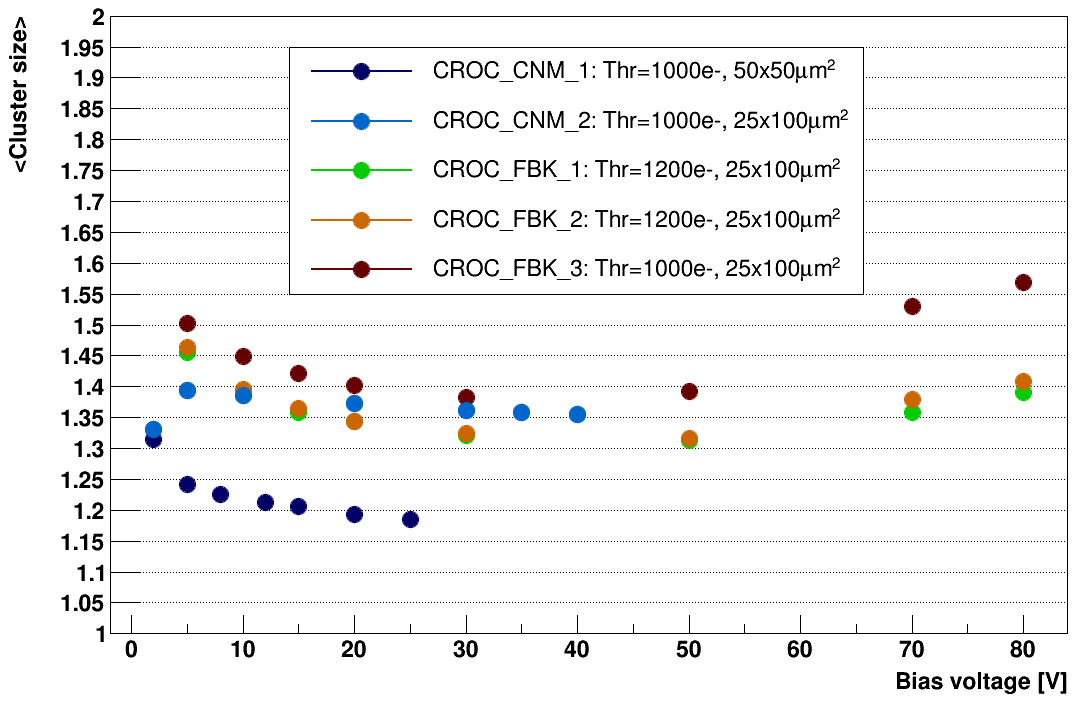}
  \label{fig:clusterSizeScan_fresh}}
\nextsubcolumn[0.4\textwidth]
  \subfloat[]{\includegraphics[width=5.9cm]{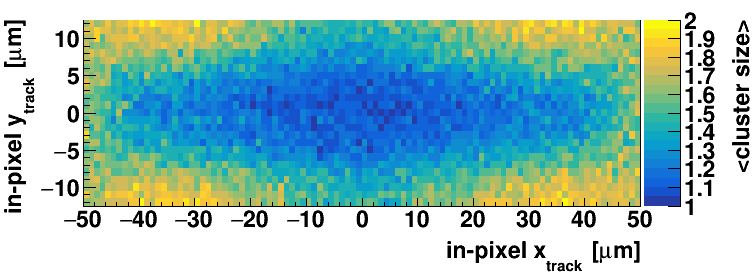}
  \label{fig:clusterSize_map_fresh5V}}
  \nextsubfigure
  \subfloat[][]{\includegraphics[width=5.9cm]{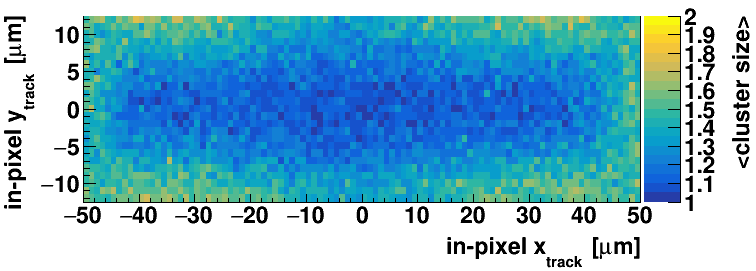}
  \label{fig:clusterSize_map_fresh30V}}
\end{subcolumns}
\vspace{-0.1cm}
\caption{(a) Average cluster size as a function of bias voltage for several non-irradiated modules. Pixel cell maps of cluster size for module CROC\_FBK\_1 at a bias voltage of (b) $\SI{5}{V}$ and (c) $\SI{30}{V}$.}
\end{figure}

The cluster size is closely related to the spatial resolution of the devices. Figure~\ref{fig:clusterSizeScan_fresh} shows the average cluster size as a function of bias voltage. The cluster size increases as the electric field extends to the periphery of the pixel cell, until the devices are fully depleted. Beyond the full depletion voltage, the electric field becomes strong enough to limit the charge sharing due to diffusion, leading to a decrease in cluster size at the periphery. This effect is clearly illustrated in the cluster size cell maps shown in Figs.~\ref{fig:clusterSize_map_fresh5V} and~\ref{fig:clusterSize_map_fresh30V}, corresponding to module CROC\_FBK\_1 at full depletion ($\SI{5}{V}$) and above full depletion ($\SI{30}{V}$), respectively. A plausible explanation for the increased cluster size at higher bias voltages is the generation of microavalanches; when the bias voltage significantly exceeds the full depletion voltage, the electric field may become strong enough to artificially enlarge the cluster size. The cluster size of the $\SI{50}{\mu m}$-pitch sensor is notably smaller because the configuration of the electric field in this cell type results in less charge sharing.

\begin{figure}[hbt!]
\centering
\begin{subcolumns}[0.4\textwidth]
  \subfloat[]{\hspace{0.1cm}\includegraphics[width=5.6cm]
{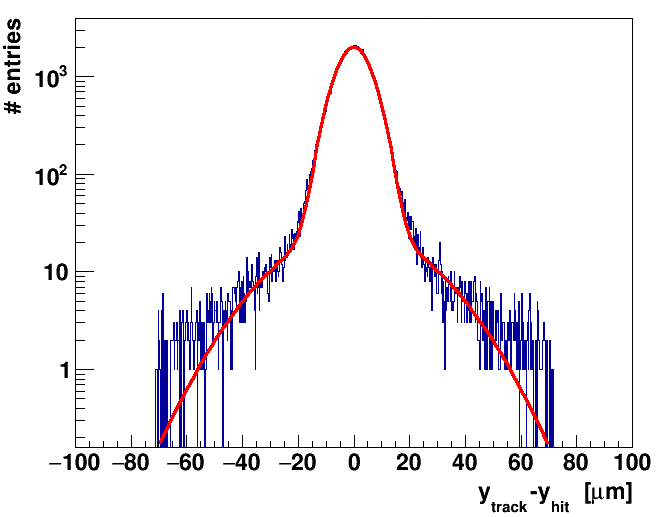}
  \label{fig:resY}}
\nextsubfigure
\subfloat[]{\includegraphics[width=5.7cm]
{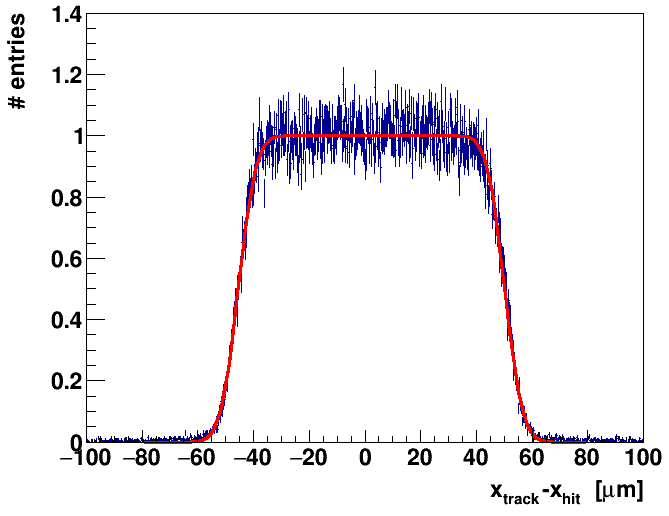}
  \label{fig:resX}}
\nextsubcolumn[0.57\textwidth]
  \subfloat[][]{\includegraphics[width=8.4cm]{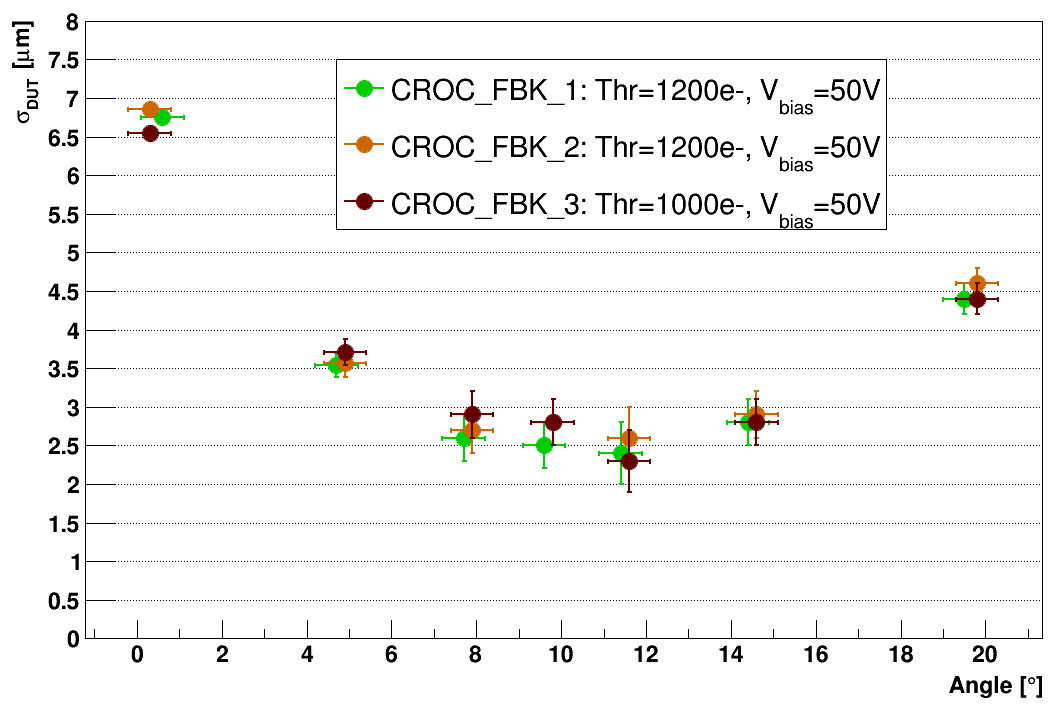}
  \label{fig:DUTresolution_fresh}}
\end{subcolumns}
\vspace{-0.2cm}
\caption{Example of the residual distribution for (a) the $\SI{25}{\mu m}$-pitch fitted to the sum of two Gaussians, and (b) the $\SI{100}{\mu m}$-pitch fitted to the difference between two cumulative distribution functions. (c) Spatial resolution on the short pixel pitch as a function of the rotation angle for several non-irradiated FBK modules biased at $\SI{50}{V}$.}
\label{fig:resolutions_fresh}
\end{figure}

The DUT spatial resolution is estimated from the residuals, defined as the differences between the track impact position on the DUT and the position of the assigned cluster. Consequently, the residual distribution is a convolution of the DUT spatial resolution and the telescope resolution. Figures~\ref{fig:resY} and~\ref{fig:resX} present examples of the residual distributions for the short ($\SI{25}{\mu m}$) and long ($\SI{100}{\mu m}$) pixel pitches, respectively. The width of the distribution for the short pixel pitch, typically obtained from a Gaussian fit, corresponds to the sum in quadrature of these two resolutions. However, this distribution exhibits tails, primarily due to misassociations between clusters and tracks. This artifact can be addressed using one of the following fitting models: the sum of two Gaussians, $N_{1}(\mu,\sigma_{1}^{2})+N_{2}(\mu,\sigma_{2}^{2})$, or the sum of a Gaussian and a constant term $c$, $N_{1}(\mu,\sigma_{1}^{2})+\textit{c}$. The mean of the standard deviations from the Gaussian fits to the core of the distribution, derived from both models, is taken as the effective width of the residual distribution in this analysis.

A data-driven approach is used to estimate the telescope resolution. Unlike on the residuals for the $\SI{25}{\mu m}$-pitch, the effects of charge sharing and telescope resolution on the residuals for the $\SI{100}{\mu m}$-pitch are disentangled, owing to low charge sharing across the short pixel pitch. Consequently, the telescope resolution is determined by fitting the edge of the residual distribution for the long pixel pitch to the cumulative distribution function of a Gaussian, $\Phi(\frac{x-\mu}{\sigma})=\frac{1}{2}(1+\text{erf}(\frac{x-\mu}{\sigma\sqrt{2}}))$. The standard deviation obtained from this fit, divided by the cosine of the angle, represents the telescope resolution at the DUT position. Figure~\ref{fig:DUTresolution_fresh} shows the spatial resolution on the short pixel pitch of the FBK modules as a function of the angle, defined as the rotation around an axis parallel to the long pixel pitch and perpendicular to the beam direction. The best resolution is about $\SI{2.5}{\mu m}$.
\vspace{-0.8cm}

\vspace{-0.8cm}
\section{Methodology and test beam results of irradiated modules}
\label{sec:irradiated_modules}

The performance after irradiation of several modules with $25\times100$~$\SI{}{\mu m^{2}}$ pixel size was also evaluated in test beams. The thresholds of the modules irradiated up to fluences of $\SI{1e16}{n_{eq}cm^{-2}}$ and $\SI{1.6e16}{n_{eq}cm^{-2}}$ were tuned at a temperature of $\SI{-30}{\degree C}$ to average pixel thresholds of $\SI{1000}{e^{-}}$ and $\SI{1200}{e^{-}}$, respectively. Pixels were masked using the same criteria described for non-irradiated devices. The FBK samples irradiated to the lowest fluence kept the percentage of masked pixels below $\SI{2}{\%}$ up to a bias voltage of $\SI{150}{V}$, whereas the CNM sample exhibited a sharp increase at $\SI{130}{V}$. At this voltage, the proportion of masked pixels in the CNM module decreased from $\SI{9}{\%}$ to $\SI{3}{\%}$ when the threshold was raised to $\SI{1200}{e^{-}}$. In the more heavily irradiated device, the percentage of masked pixels remained below $\SI{3}{\%}$ up to $\SI{130}{V}$, but then rose steeply to $\SI{10}{\%}$ at $\SI{140}{V}$.

\begin{figure}[hbt!]
\centering
\begin{subcolumns}[0.1\textwidth]
  \subfloat[]{\includegraphics[width=7.4cm]{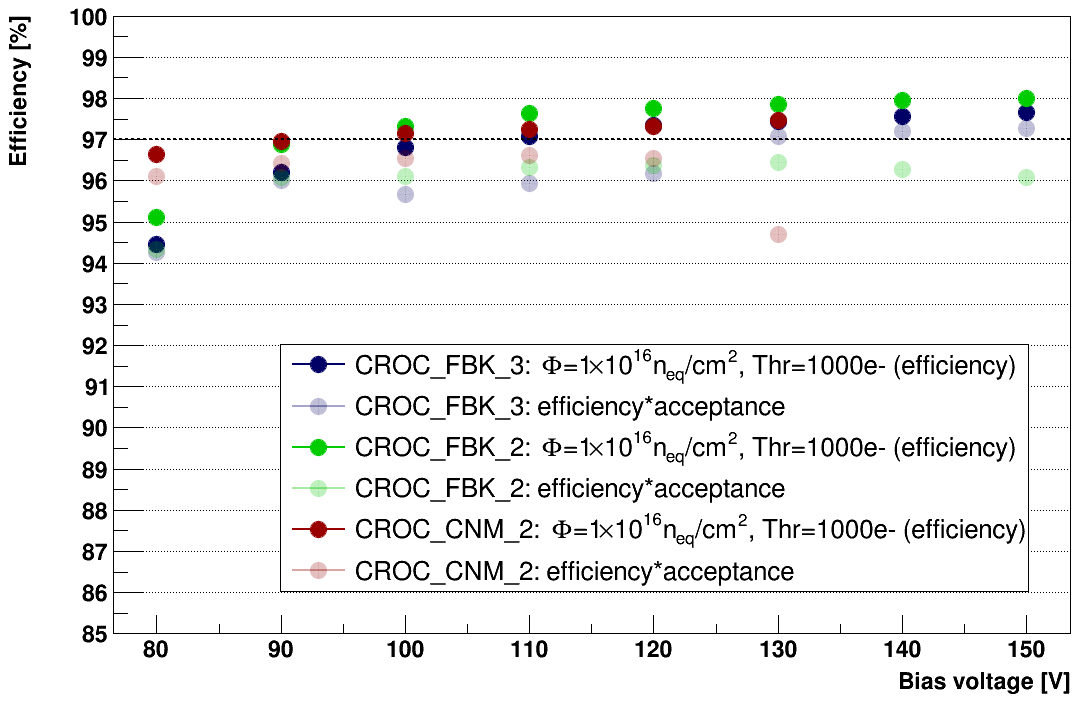}
  \label{fig:effxAccScan_irrad}}
\nextsubcolumn[0.49\textwidth]
  \subfloat[]{\includegraphics[width=7.4cm]{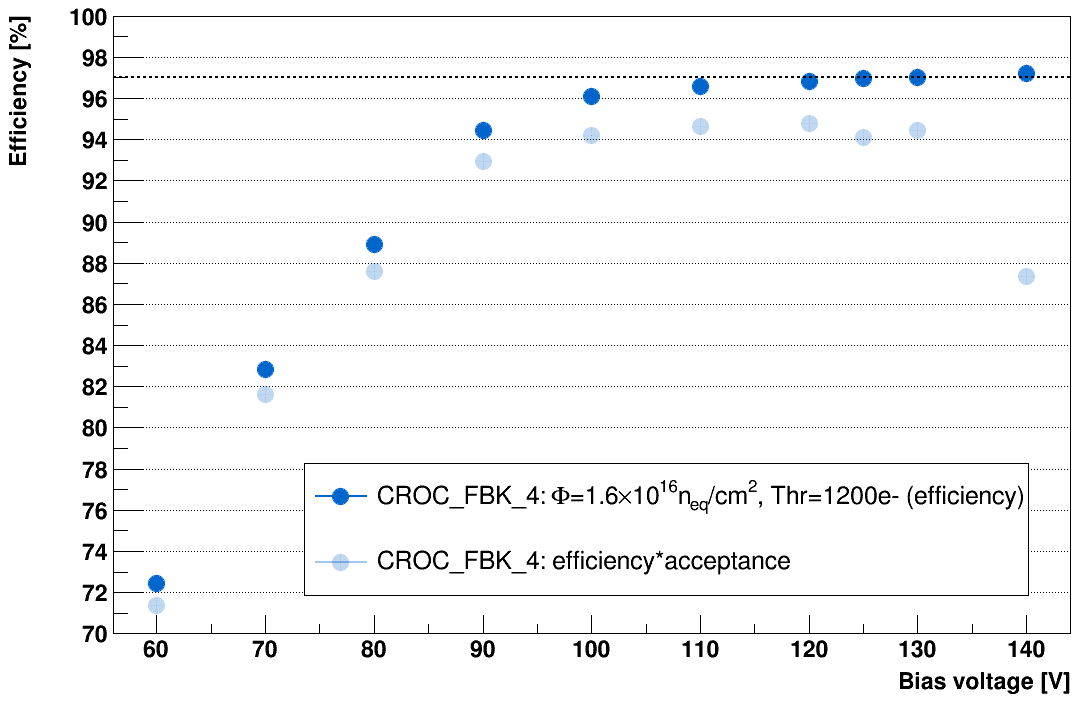}
  \label{fig:effxAccScan_irrad1d6e16}}
\end{subcolumns}
\vspace{-0.2cm}
\caption{Hit detection efficiency with and without the acceptance correction as a function of the bias voltage for modules irradiated to (a) $\SI{1e16}{n_{eq} cm^{-2}}$ and (b) $\SI{1.6e16}{n_{eq} cm^{-2}}$. The solid colors show the efficiency, while the lighter colors represent the efficiency times the acceptance.}
\label{fig:effxAcc}
\end{figure}

Figures~\ref{fig:effxAccScan_irrad} and~\ref{fig:effxAccScan_irrad1d6e16} show the hit detection efficiency at normal beam incidence as a function of the bias voltage for modules irradiated to $\SI{1e16}{n_{eq} cm^{-2}}$ and $\SI{1.6e16}{n_{eq} cm^{-2}}$, respectively. $\SI{97}{\%}$ efficiency is achieved by all modules, resulting in a wide operation range with a low number of masked pixels and excellent performance: around $\SI{50}{V}$ for modules irradiated to $\SI{1e16}{n_{eq} cm^{-2}}$ and $\SI{30}{V}$ for those irradiated to $\SI{1.6e16}{n_{eq} cm^{-2}}$. The efficiency corrected by the acceptance, which is defined as $1-\frac{\text{No. of masked pixels}}{\text{No. of pixels}}$, is also illustrated in Fig.~\ref{fig:effxAcc} to quantify the impact of masked pixels on the detection efficiency. Whenever the percentage of masked pixels sharply increases, the corrected efficiency significantly drops, as seen in Fig.~\ref{fig:effxAccScan_irrad1d6e16} at the highest bias voltage.

\begin{figure}[hbt!]
\centering
\begin{subcolumns}[0.5\textwidth]
  \subfloat[]{\includegraphics[width=7.4cm]{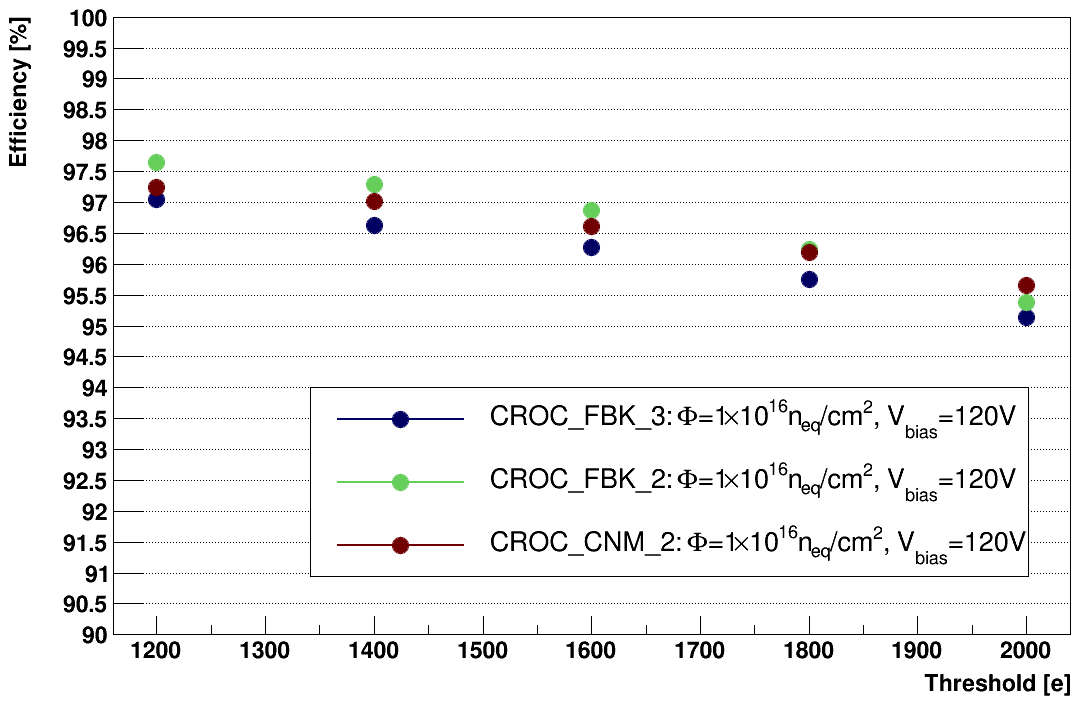}
  \label{fig:thresholdScan_irrad}}
\nextsubcolumn[0.5\textwidth]
  \subfloat[]{\includegraphics[width=7.4cm]{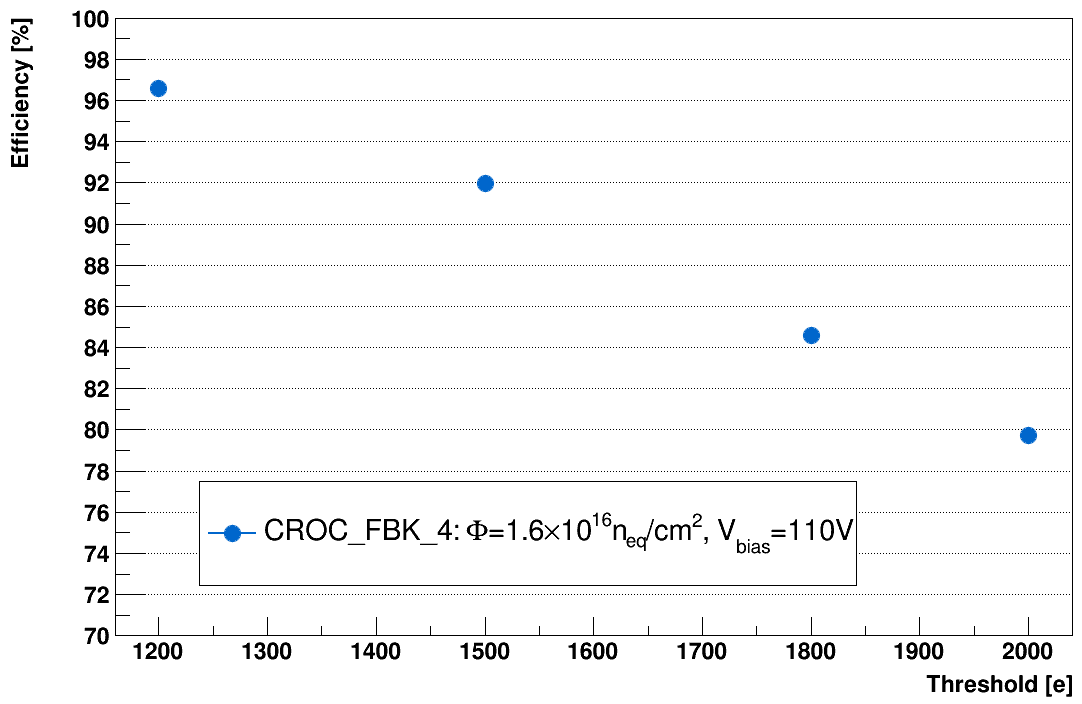}
  \label{fig:thresholdScan_irrad1e16}}
\end{subcolumns}
\caption{Hit detection efficiency as a function of the average pixel threshold for (a) modules irradiated to $\SI{1e16}{n_{eq}cm^{-2}}$ at $\SI{120}{V}$ and (b) modules irradiated to $\SI{1.6e16}{n_{eq}cm^{-2}}$ at $\SI{110}{V}$.}
\label{fig:thresholdScan}
\end{figure}

The dependence of the efficiency on the average pixel threshold to which the modules are tuned is shown in Fig.~\ref{fig:thresholdScan} for both fluences. This study was conducted at full depletion, normal beam incidence and without a pixel mask. The maximum efficiency drop observed in the less irradiated modules is $\SI{2}{\%}$ when doubling the threshold, while it becomes $\SI{16}{\%}$ for the heavily irradiated one. These results underline the importance of keeping the threshold as low as possible at high fluences.

\begin{figure}[hbt!]
\centering
\begin{subcolumns}[0.1\textwidth]
  \subfloat[]{\includegraphics[width=6.8cm]{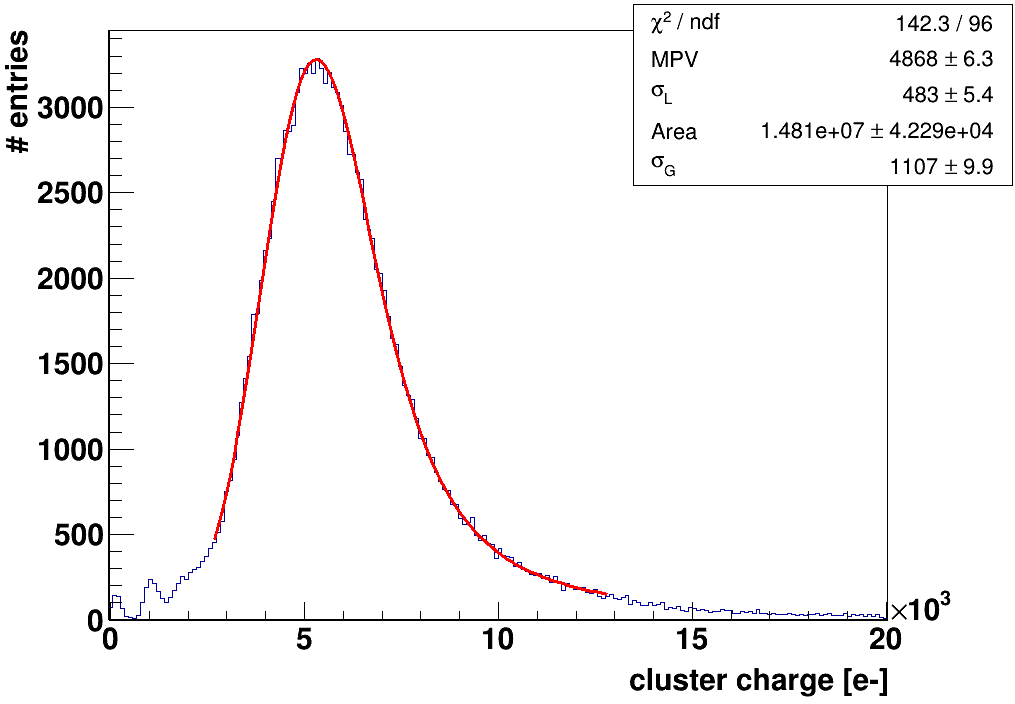}
  \label{fig:clusterCharge_irrad}}
\nextsubcolumn[0.49\textwidth]\vspace{-0.4cm}
  \subfloat[]{\includegraphics[width=7.4cm]{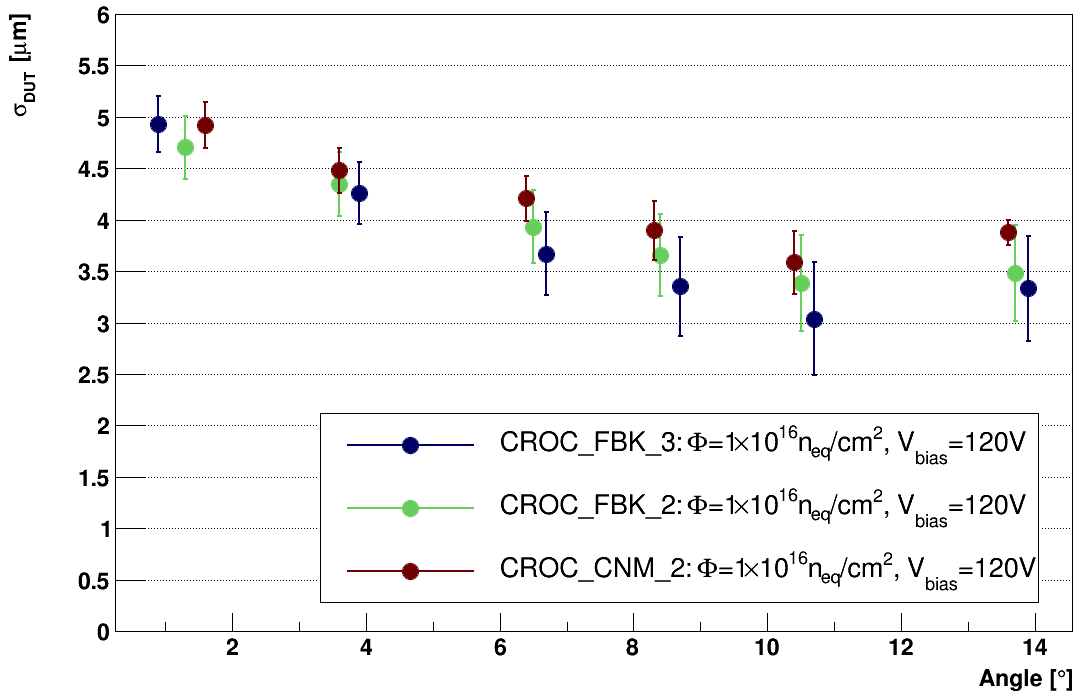}
  \label{fig:resolution_irrad}}
\end{subcolumns}
\caption{(a) Cluster charge distribution for module CROC\_FBK\_2 irradiated to $\SI{1e16}{n_{eq}\cm^{-2}}$ and biased at $\SI{130}{V}$. It is fitted to a Landau distribution with most probable value $MPV$ and width $\sigma_{L}$, convoluted with a Gaussian of width $\sigma_{G}$. (b) Spatial resolution on the short pixel pitch as a function of the rotation angle for modules irradiated to $\SI{1e16}{n_{eq}\cdot cm^{-2}}$ and biased at $\SI{120}{V}$.}
\end{figure}

Irradiation leads to a degradation in both the amount of collected charge and the spatial resolution of the devices. Figure~\ref{fig:clusterCharge_irrad} shows the cluster charge distribution at normal beam incidence for module CROC\_FBK\_2, irradiated to $\SI{1e16}{n_{eq}cm^{-2}}$ and biased at $\SI{130}{V}$. The cluster charge MPV is around $\SI{5000}{e-}$, indicating a $\SI{50}{\%}$ reduction compared to the charge expected before irradiation. For the short pixel pitch, the best spatial resolution after irradiation to $\SI{1e16}{n_{eq}cm^{-2}}$ is approximately $\SI{3.5}{\mu m}$, as shown in Fig.~\ref{fig:resolution_irrad}.

\vspace{-0.2cm}
\section{Conclusions}
\label{sec:conclusions}
\vspace{-0.2cm}

3D pixel sensors equipped with the prototype of the final readout chip have proven to have excellent performance, meeting all CMS requirements. After irradiation up to $\SI{1.6e16}{n_{eq}cm^{-2}}$, these devices maintain a hit detection efficiency above $\SI{97}{\%}$ across a wide operation range, with a low average pixel threshold and a small number of masked pixels.

\bibliography{iWoRiD2024_CLG}

\end{document}